\documentclass[aps,prl,twocolumn,superscriptaddress,amsmath,amssymb,footinbib,longbibliography,showkeys]{revtex4-1}
\usepackage[english]{babel}
\usepackage{amssymb,amsbsy,amsmath}
\usepackage{graphicx}
\usepackage{dcolumn}
\usepackage{bm}
\usepackage{subfigure}
\usepackage{color}
\usepackage{textcomp}
\usepackage[colorlinks,bookmarks=false,citecolor=blue,linkcolor=red,urlcolor=blue]{hyperref}

\newcommand{\op}[1]{\fontdimen12\textfont3=2pt\fontdimen12\scriptfont3=1.4pt\!\null\mathop{\protect\vphantom{#1}\smash{#1}}\limits_{\sim}\null\!}
\newcommand{\EinsOp}
           {\;\smash{\raisebox{-1.1ex}{$\!\!\stackrel{\!\mbox{1}
            \hspace{-0.4ex}\rule[0.0ex]{0.06ex}{1.60ex}}{\sim}$}}}
\newcommand{\xref}[1]{\protect\ref{#1}}
\newcommand{\figref}[1]{Fig.~\protect\ref{#1}}
\newcommand{\fmref}[1]{(\protect\ref{#1})}

\def\ket#1{\, | \, {#1} \, \rangle}

\begin{document}
\title{Supersymmetric spin-phonon coupling prevents odd integer spins from quantum tunneling}

\author{Kilian Irl\"ander}
\affiliation{Fakult\"at f\"ur Physik, Universit\"at Bielefeld, Postfach 100131, D-33501 Bielefeld, Germany}
\author{Heinz-J\"urgen Schmidt}
\affiliation{Universit\"at Osnabr\"uck, Fachbereich Physik,
 	D-49069 Osnabr\"uck, Germany}
\author{J\"urgen Schnack}
\email{jschnack@uni-bielefeld.de}
\affiliation{Fakult\"at f\"ur Physik, Universit\"at Bielefeld, Postfach 100131, D-33501 Bielefeld, Germany}

\date{\today}

\begin{abstract}
Quantum tunneling of the magnetization is a phenomenon that
impedes the use of small anisotropic spin systems for storage
purposes even at the lowest temperatures. Phonons, usually
considered for temperature dependent relaxation of magnetization
over the anisotropy barrier, also contribute to magnetization
tunneling for integer spin quantum numbers. Here we demonstrate
that certain spin-phonon Hamiltonians are unexpectedly robust
against the opening of a tunneling gap, even for strong
spin-phonon coupling. The key to understanding
this phenomenon is provided by an underlying
supersymmetry that involves both spin and phonon degrees of freedom.
\end{abstract}


\maketitle

\begin{figure}[ht!]
\centering
\includegraphics*[clip,width=0.45\columnwidth]{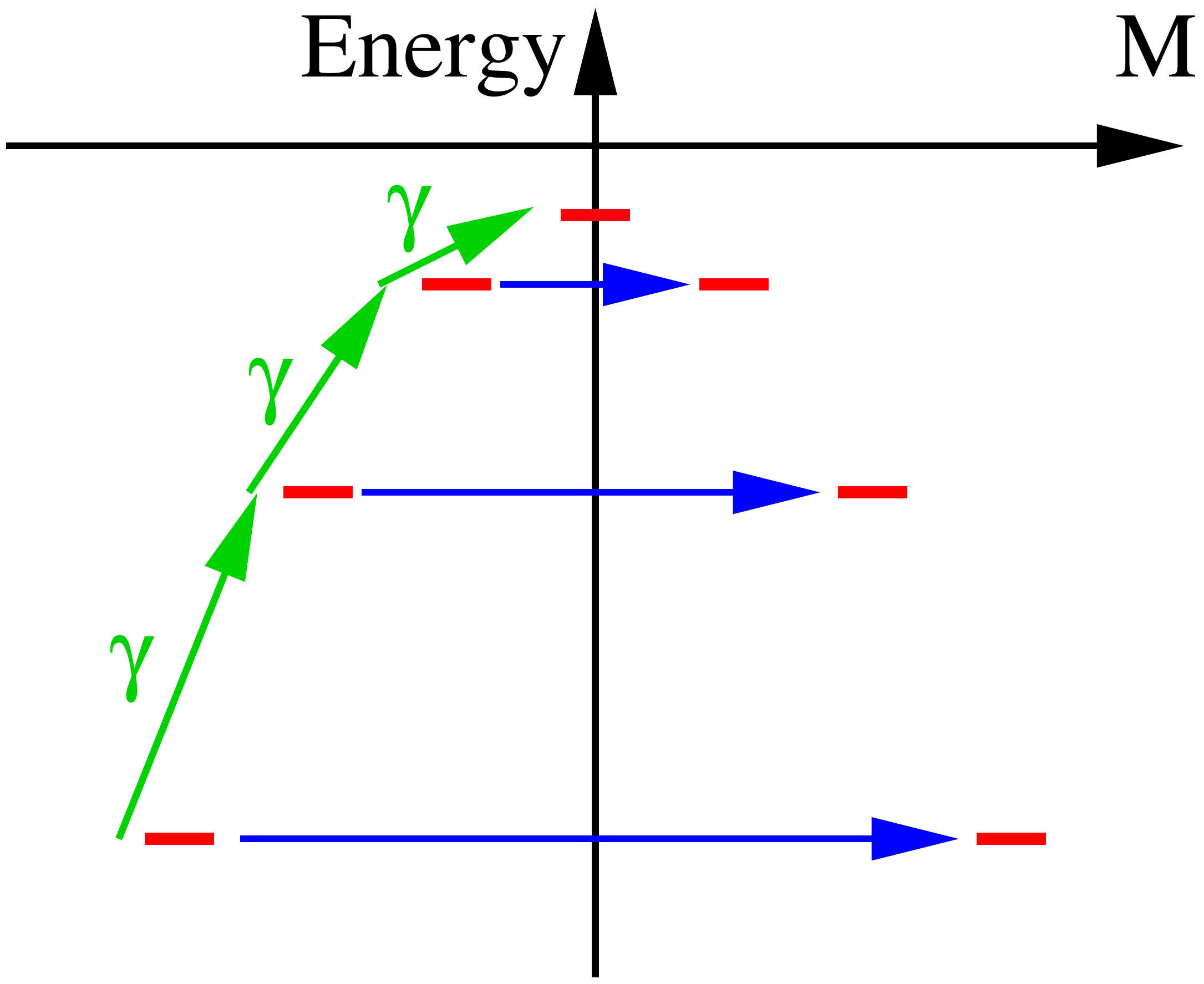}
\quad
\includegraphics*[clip,width=0.40\columnwidth]{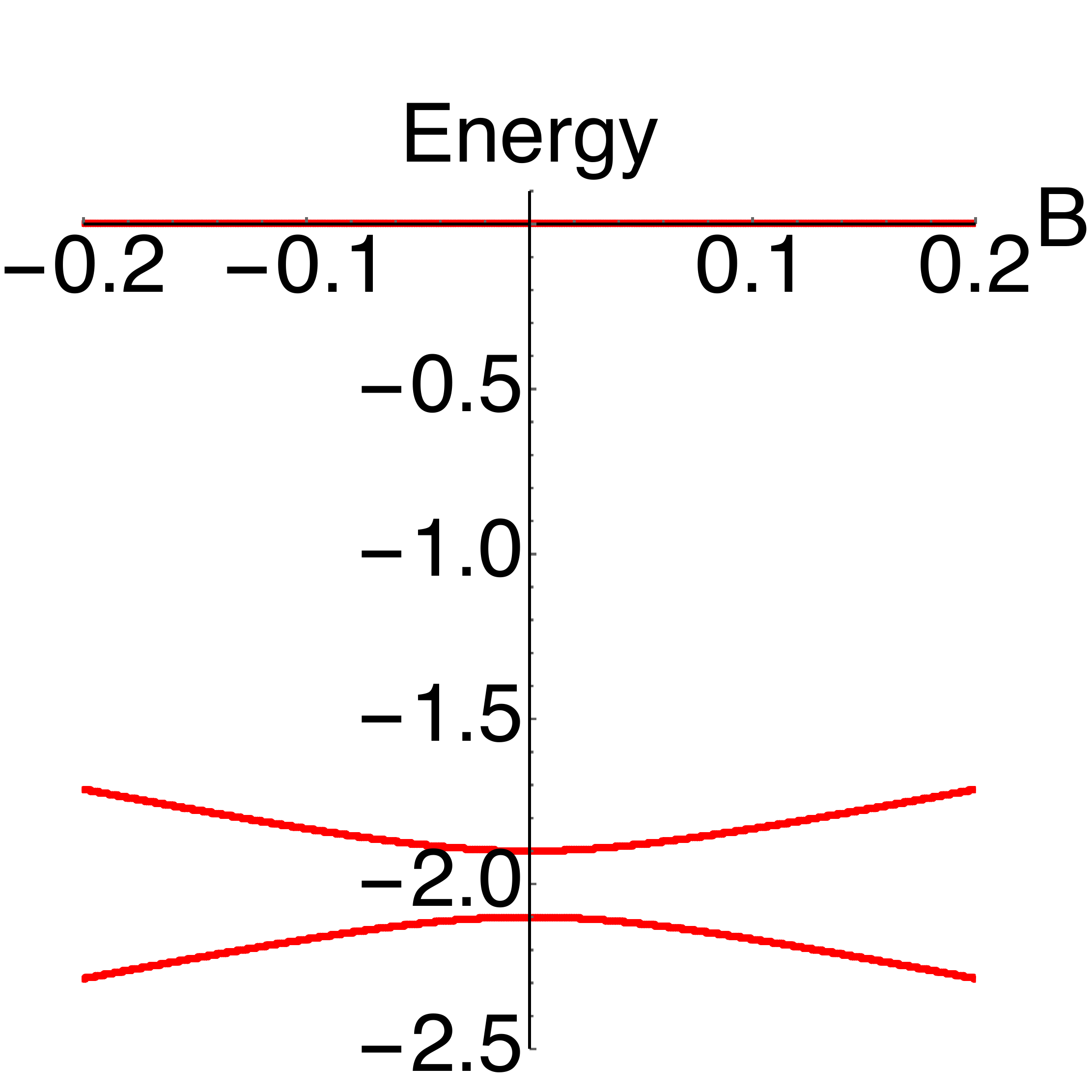}
\caption{L.h.s.: Sketch of the low-lying energy levels of a spin
  $s=3$ with dominant easy-axis anisotropy
  vs.\ magnetic quantum number.
  Red bars denote energy eigenvalues. Blue arrows show
  magnetization tunneling pathways for states with negative
  magnetic quantum number, and green arrows depict some of the
  possible excitations due to phonons, compare
  e.g.\ \cite{LvS:CSR15}.
  R.h.s.: Example of a tunnel splitting for a spin
  $s=1$ with $D<0$ and $E\neq 0$ anisotropy terms.} 
\label{spin-phonon-susy-f-A}
\end{figure}

\emph{Introduction.}---Single-ion magnetic anisotropy provides
the simplest mechanism for fundamental phenomena such as
magnetic bistability as well as quantum tunneling of the
magnetization \cite{GSV:2006,Blundell:CP07,Sch:CP19}. The Hamiltonian is so simple that any student
after an introductory course on quantum mechanics can
diagonalize it. It consists of two terms,
\begin{eqnarray}
  \op{H}_{\text{SI}}
&=&
  D (\op{s}_z)^2 + E \left\{ (\op{s}_x)^2 -(\op{s}_y)^2 \right\}
  \label{E-000}
\\
  &=&
  D (\op{s}_z)^2 + \frac{E}{2} \left\{ (\op{s}^+)^2 +(\op{s}^-)^2 \right\}
  \label{E-000B}
  \ ,
\end{eqnarray}
which, for obvious reasons, have been termed $D$- and $E$-term,
see e.g. \cite{GSV:2006} for a full account of the story.
A negative $D$, $D<0$, results in an easy-axis anisotropy which,
in the absence of the $E$-term,
would express itself as a perfect parabolic anisotropy barrier,
compare l.h.s.\ of \figref{spin-phonon-susy-f-A}. $E$ leads to a
splitting of the otherwise degenerate pairs of states left and
right of the barrier if the considered spin is integer,
compare r.h.s.\ of \figref{spin-phonon-susy-f-A}. If the spin is
half integer, Kramers' theorem applies, and the levels are
bound to be degenerate for $B=0$.

In a magnetic field applied along the easy-axis one encounters a
perfect level crossing for $E=0$; such systems - single ion
magnets (SIM) or single molecule magnets (SMM) -- do show
bistability of the magnetization and are thus suitable
candidates for magnetic storage devices
\cite{SGC:Nat93,FST:PRL96,TLB:Nature96,LTB:JAP97,CWM:PRL00,WSC:JAP00,GaS:ACIE03,Goo:DT20}. In
case of a splitting of the two 
lowest levels, one observes an avoided level crossing as a
function of applied field as depicted
on the r.h.s.\ of \figref{spin-phonon-susy-f-A}. The
magnetization is not bi-stable at $B=0$, instead it tunnels
as described for two-level systems by Landau, Zener, and
Stueckelberg
\cite{Zen:PRS32,Lan:PZS32,Stueckelberg:HPA32}. The splitting
therefore is also called tunnel splitting. 

\begin{figure}[t]
\centering
\includegraphics*[clip,width=0.80\columnwidth]{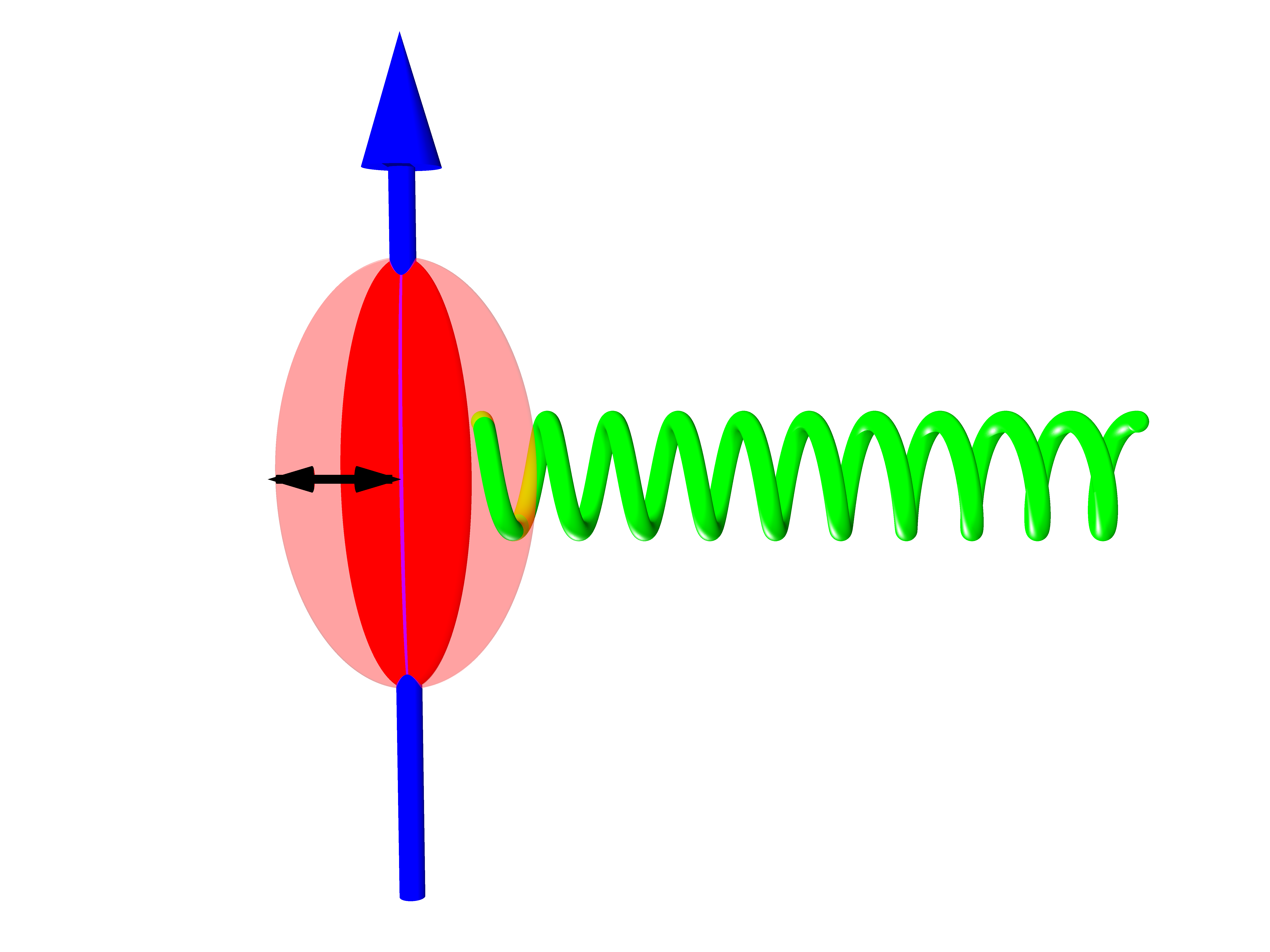}
\caption{Sketch of the coupling of the anisotropy tensor to
  phonons of the material. The reddish ellipsoid represents the
  anisotropy tensor whose components ($E$-terms) along major
  axes perpendicular
  to the easy axis are modified via a coupling to a special
  phonon mode (green coil). For a relation to the strain tensor compare
  e.g.\ the first term of Eq.~(6) of Ref.~\cite{LeL:PRB00} for a
  specific example.} 
\label{spin-phonon-susy-1}
\end{figure}

To our surprise this simple scheme -- tunnel splitting for
integer spins, no tunnel splitting for half-integer spins --
needs a modification for integer spins in the case of
phonon-induced tunnel splitting, if the spin is coupled
to phonons of the material in a certain way. It may then happen
that the tunnel splitting opens up only for even spin quantum
numbers, whereas one observes a perfect level crossing in the
case of odd spin quantum numbers.

This behavior is exceptional for two reasons. The insight that
spin-phonon interactions open a tunnel splitting dates back to
the late 1960's \cite{Shr:PR69,Shr:PLA70,Shr:CPL73} and was
discussed in connection with molecular magnetism ever since
then, see
e.g. \cite{VMB:PSS78,PBK:PRL02,RBM:NC16,ESG:JPCL17,MST:NC18,ORD:DT19,IrS:PRB20}.
Here we demonstrate with an example that phonon modes exist that
do not open up a tunneling gap, independent of the spin-phonon
coupling strength. The second unexpected finding is that this
behavior is not rooted in the phonon subsystem alone but can be traced
back to a combined symmetry of the spin and phonon modes, which
resembles a supersymmetry of the problem.
Note that the authors of \cite{NeW:PZ29} have already recognized
that the rules concerning the occurrence of avoided level
crossings are overridden by existing symmetries. We provide a
fundamental example.

\emph{Method.}---Specifically, we consider the
following Hamiltonian
\begin{eqnarray}
  \op{H}
  &=&
  \op{H}_{\text{SI}}
  +
  \op{H}_{\text{HO}}
  +
  \op{H}_{\text{Zeeman}}
  \ ,
  \label{E-001}
\end{eqnarray}
where the interaction of the spin with the phonons of the system is
reduced to a single harmonic oscillator,
\begin{eqnarray}
  \op{H}_{\text{HO}}
  &=&
\omega \left(\op{a}^{\dagger}\op{a}^{\mathstrut}+{\textstyle\frac{1}{2}}\right)
  \ ,
  \label{E-002}
\end{eqnarray}
for educational reasons. $\op{a}^{\dagger}$ and $\op{a}$ are the
creation and destruction operators of a certain normal mode that
couples to the spin as outlined below. The spin also 
interacts with the external magnetic field along the easy axis
described by $\op{H}_{\text{Zeeman}}$.

Key to our observation is the way the oscillator mode couples to the
spin. Out of the many couplings possible
\cite{LeL:PRB00,PBK:PRL02,CGS:PRB05,RBM:NC16,ESG:JPCL17,MST:NC18,ORD:DT19},
we investigate those cases where the phonons modify only the
$E$-terms, compare \figref{spin-phonon-susy-1} and first term of
Eq.~(6) of Ref.~\cite{LeL:PRB00} for a specific relation to the
strain tensor. We assume two different couplings, a linear coupling 
\begin{align}
  E=\alpha_1\left(\op{a}^{\dagger}+\op{a}\right)
  \ ,
\end{align}
where $E$ is proportional to the generalized coordinate of the
normal mode as well as a quadratic coupling
\begin{align}
  E=\alpha_2\left(\op{a}^{\dagger}+\op{a}\right)^2
  \ .
\end{align}
It will later turn out that the fundamental difference we found
exists between odd and even powers of the generalized coordinate
$\left(\op{a}^{\dagger}+\op{a}\right)$
of the normal mode.

Hamiltonian \fmref{E-001} can be diagonalized numerically
exactly using the product basis $\ket{m, n}$, with $m$ being the
magnetic quantum number and $n$ the oscillator quantum number,
if $n$ is cut at some maximal value $n_{\text{max}}$. We
investigated various $n_{\text{max}}=0,1,\dots 5$, and it turns
out that small $n_{\text{max}}$, even $n_{\text{max}}=1$, are
sufficient to accurately describe ground state properties
\cite{IrS:PRB20}.

\begin{figure}[ht!]
\centering
\includegraphics[width=0.45\columnwidth]{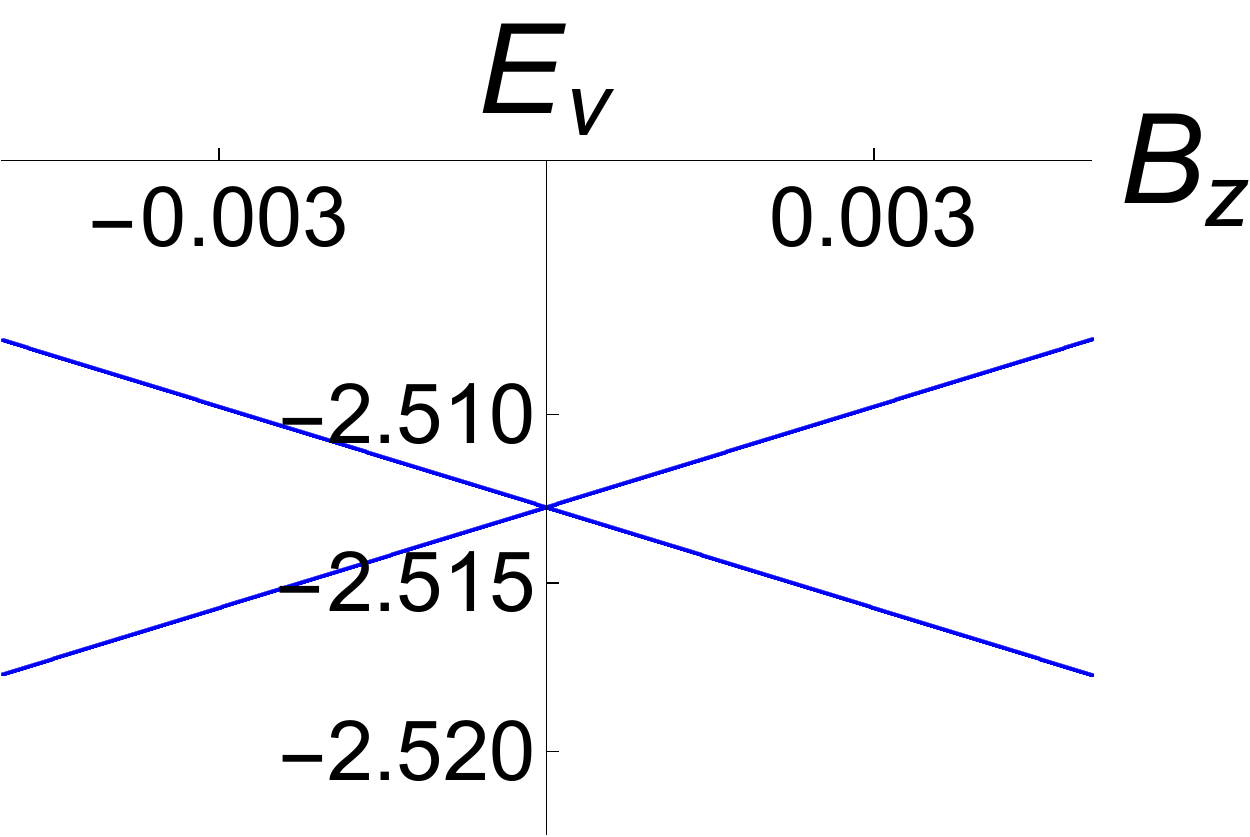}\quad
\includegraphics[width=0.45\columnwidth]{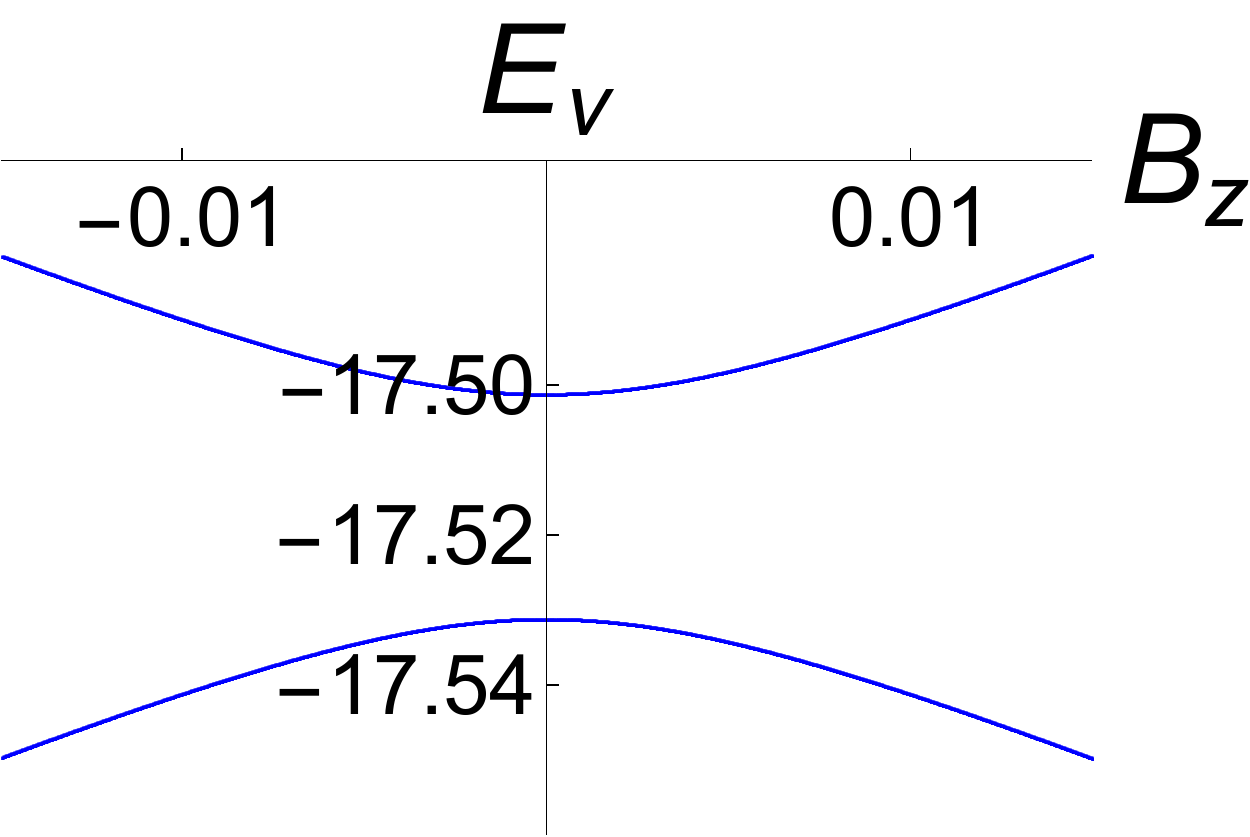}\\
$\ $\\
\includegraphics[width=0.45\columnwidth]{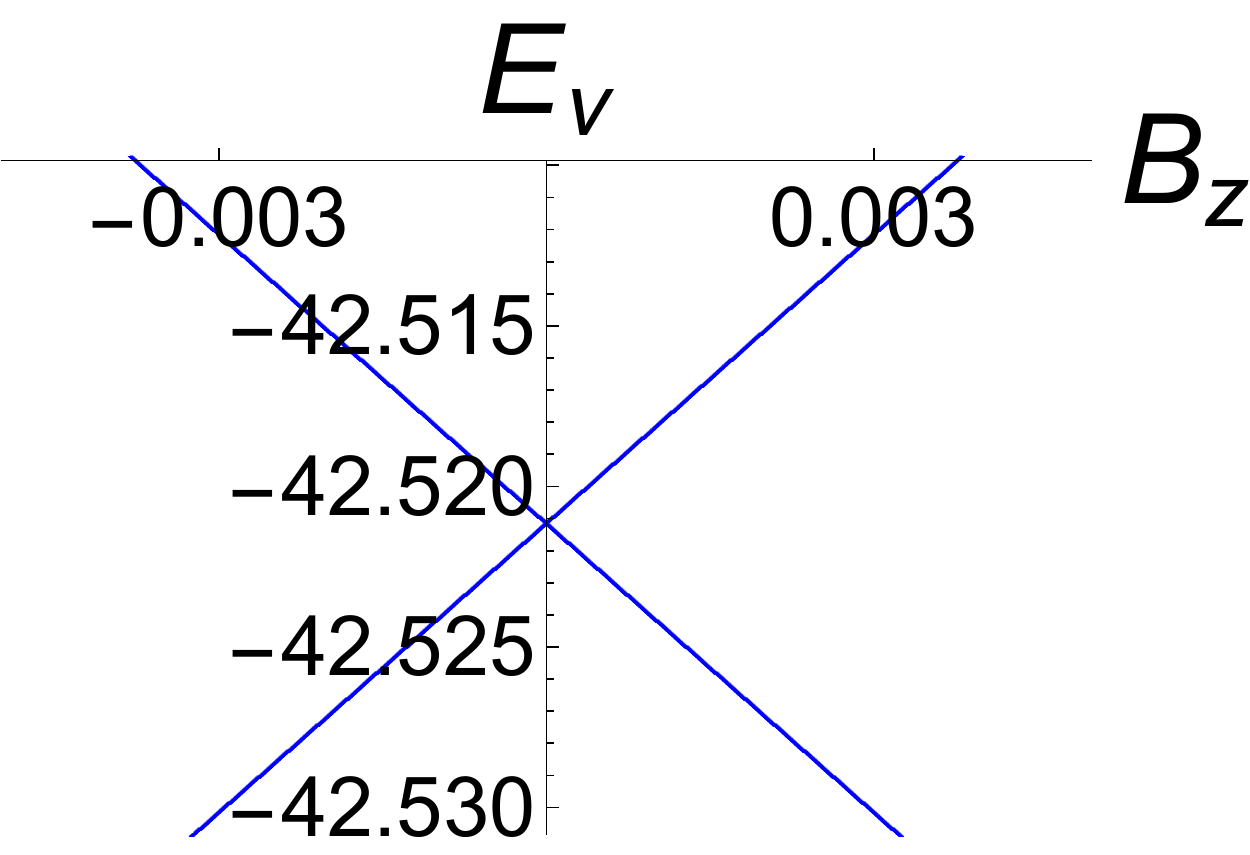}\quad
\includegraphics[width=0.45\columnwidth]{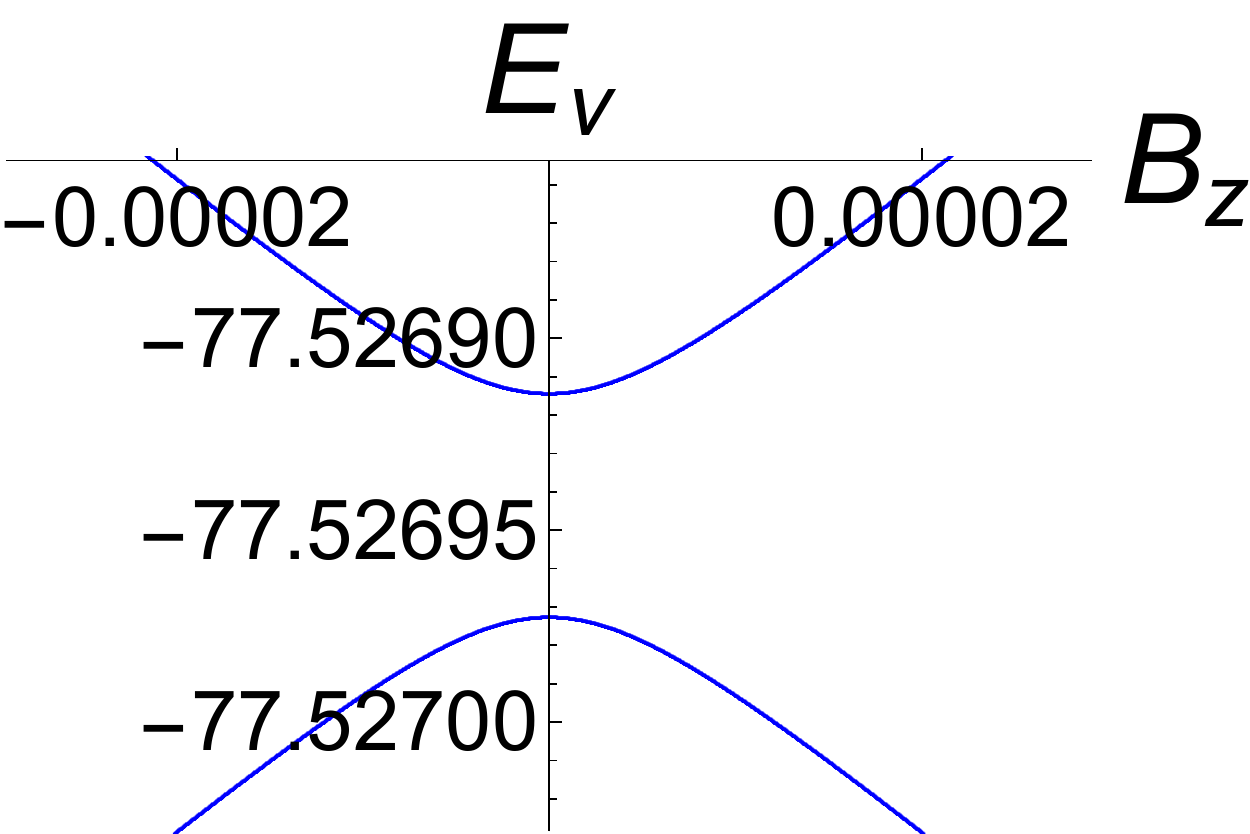}
\caption{Linear coupling: Lowest energy eigenvalues vs. magnetic
  field strength for different integer spin quantum numbers
  $s=\{1,...,4\}$ (from left to right and top to bottom) with
  $D=-5$, $n_{\text{max}}=1$, 
  $\alpha_{1}=0.5$, $\omega=5$ in natural units.} 
\label{spin-phonon-susy-2}
\end{figure}
 
\begin{figure}[ht!]
\centering
\includegraphics[width=0.45\columnwidth]{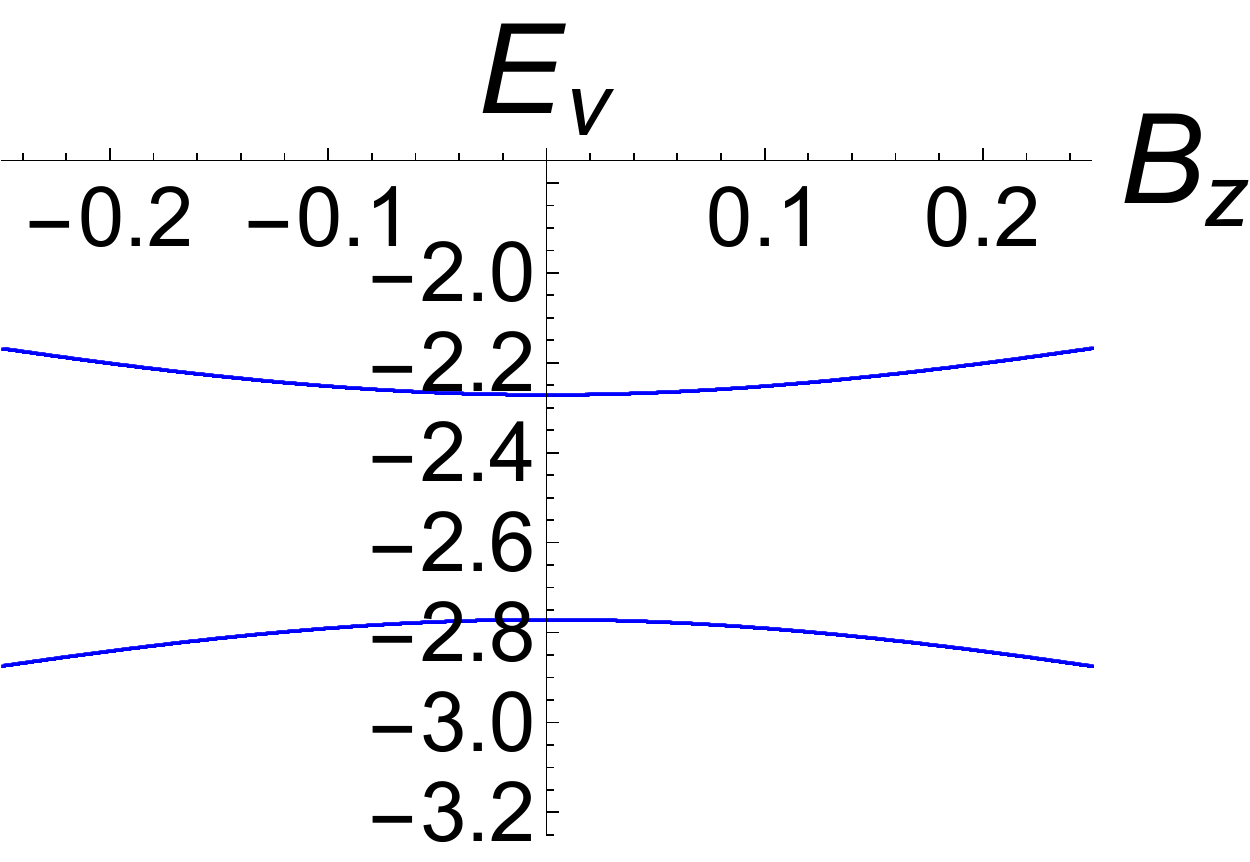}\quad
\includegraphics[width=0.45\columnwidth]{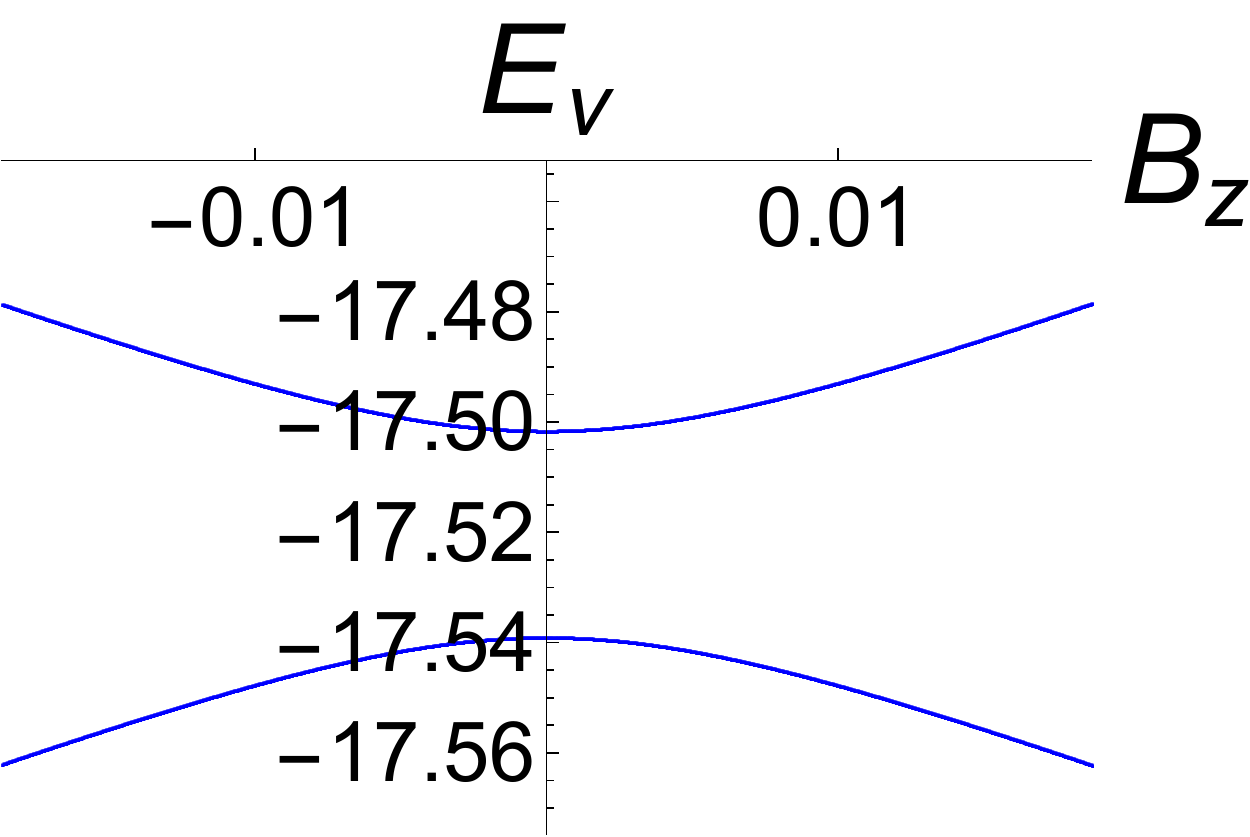}\\
$\ $\\
\includegraphics[width=0.45\columnwidth]{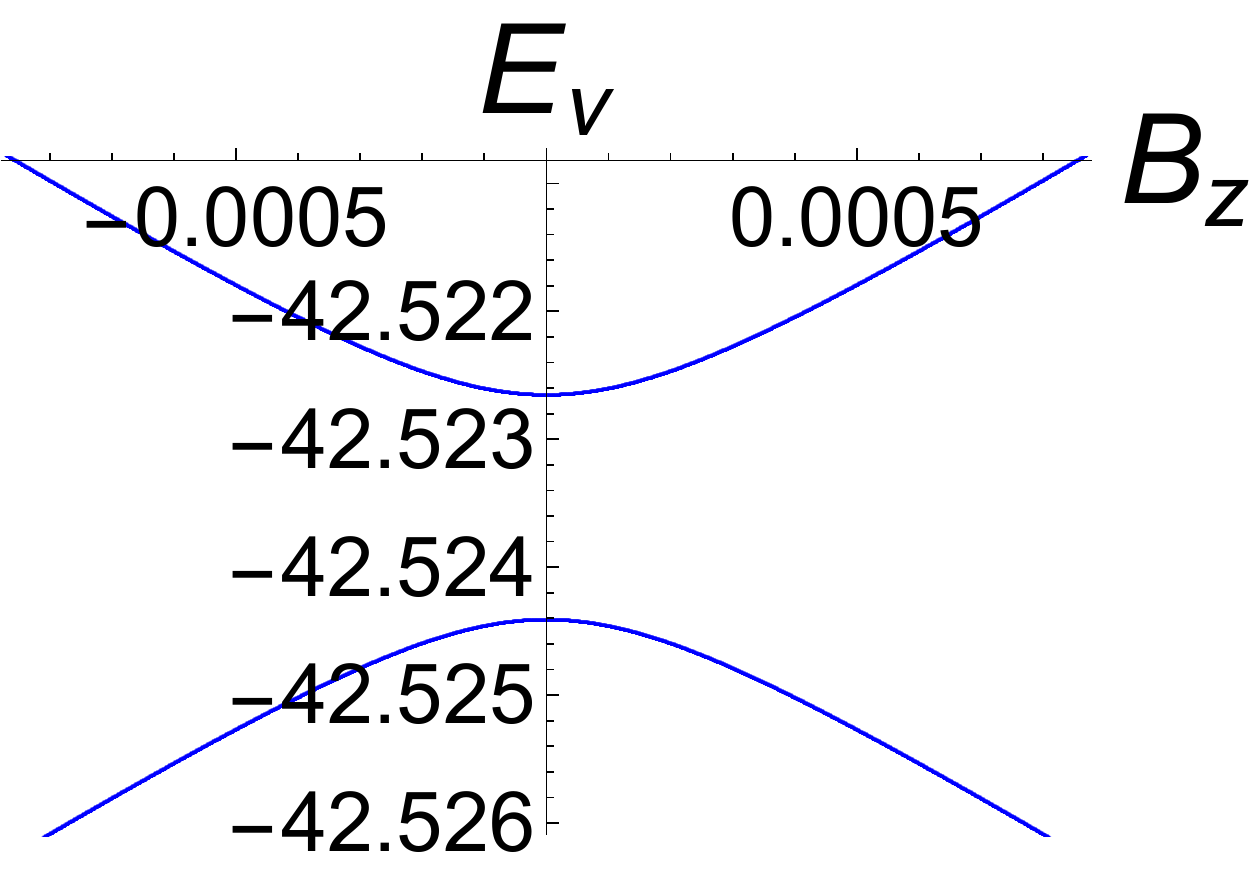}\quad
\includegraphics[width=0.45\columnwidth]{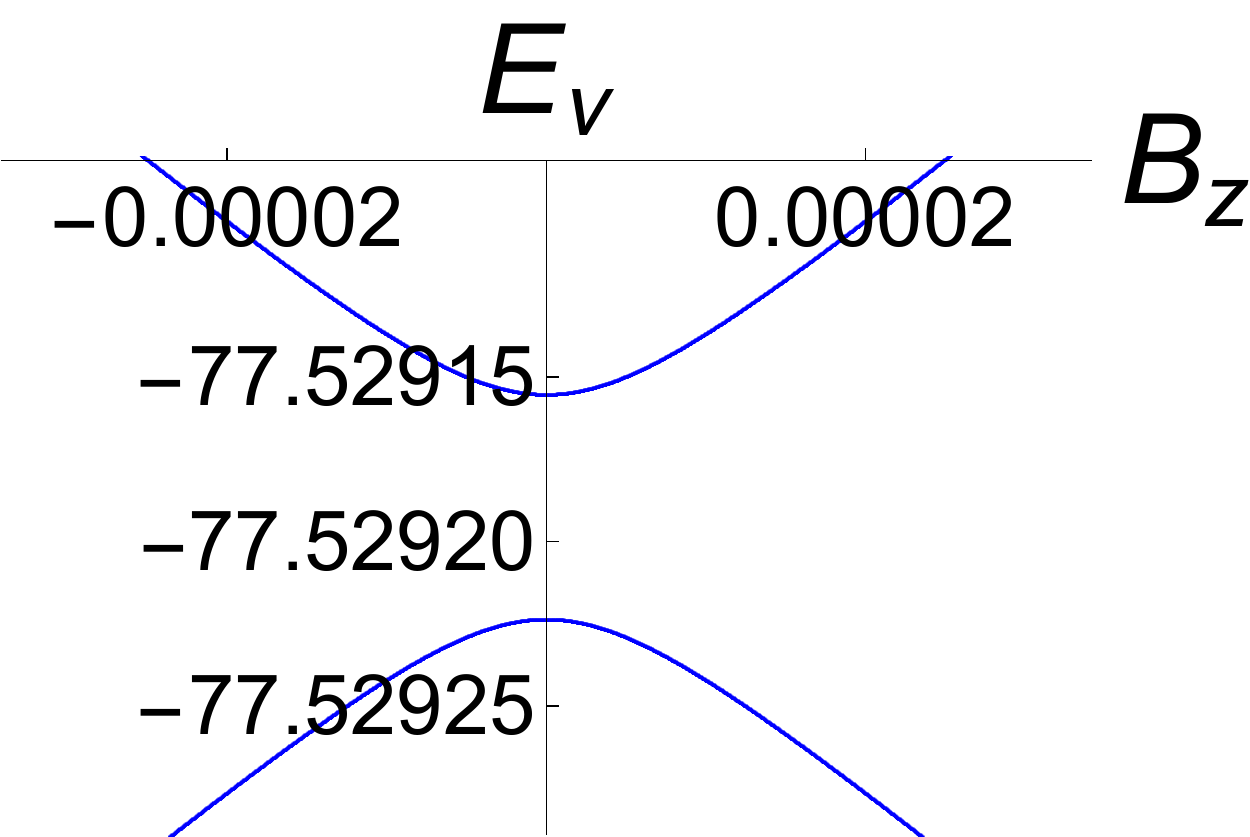}
\caption{Quadratic coupling: Lowest energy values vs. magnetic
  field strength for different integer spin quantum numbers
  $s=\{1,...,4\}$ (from left to right and top to bottom) with
  $D=-5$, $n_{\text{max}}=1$, 
  $\alpha_{2}=0.5$, $\omega=5$ in natural units.} 
\label{spin-phonon-susy-3}
\end{figure}

\emph{Results.}---A numerical diagonalization with practically
arbitrary parameters reveals that an odd-even effect determines
the tunnel splitting for the linear coupling, see
\figref{spin-phonon-susy-2}. For the quadratic coupling, the
tunneling gap opens for all integer spins, compare
\figref{spin-phonon-susy-3}. The behavior persists for higher
spin quantum numbers and $n_{\text{max}}$ as we have numerically
verified.

The case of a \emph{quadratic coupling}, or in general of a coupling
with an even power of $\left(\op{a}^{\dagger}+\op{a}\right)$, can
be immediately understood when considering that, whatever the
eigenstates of the total Hamiltonian \fmref{E-001}, the
oscillator part will contribute zero-point motion,
i.e. a parameter $E$ definitely larger than zero. Similar to
the case with constant $E$, these yield a tunnel splitting for
all integer spin quantum numbers \cite{ORD:DT19,IrS:PRB20}.

\begin{figure}[ht!]
\centering
\includegraphics*[clip,width=0.80\columnwidth]{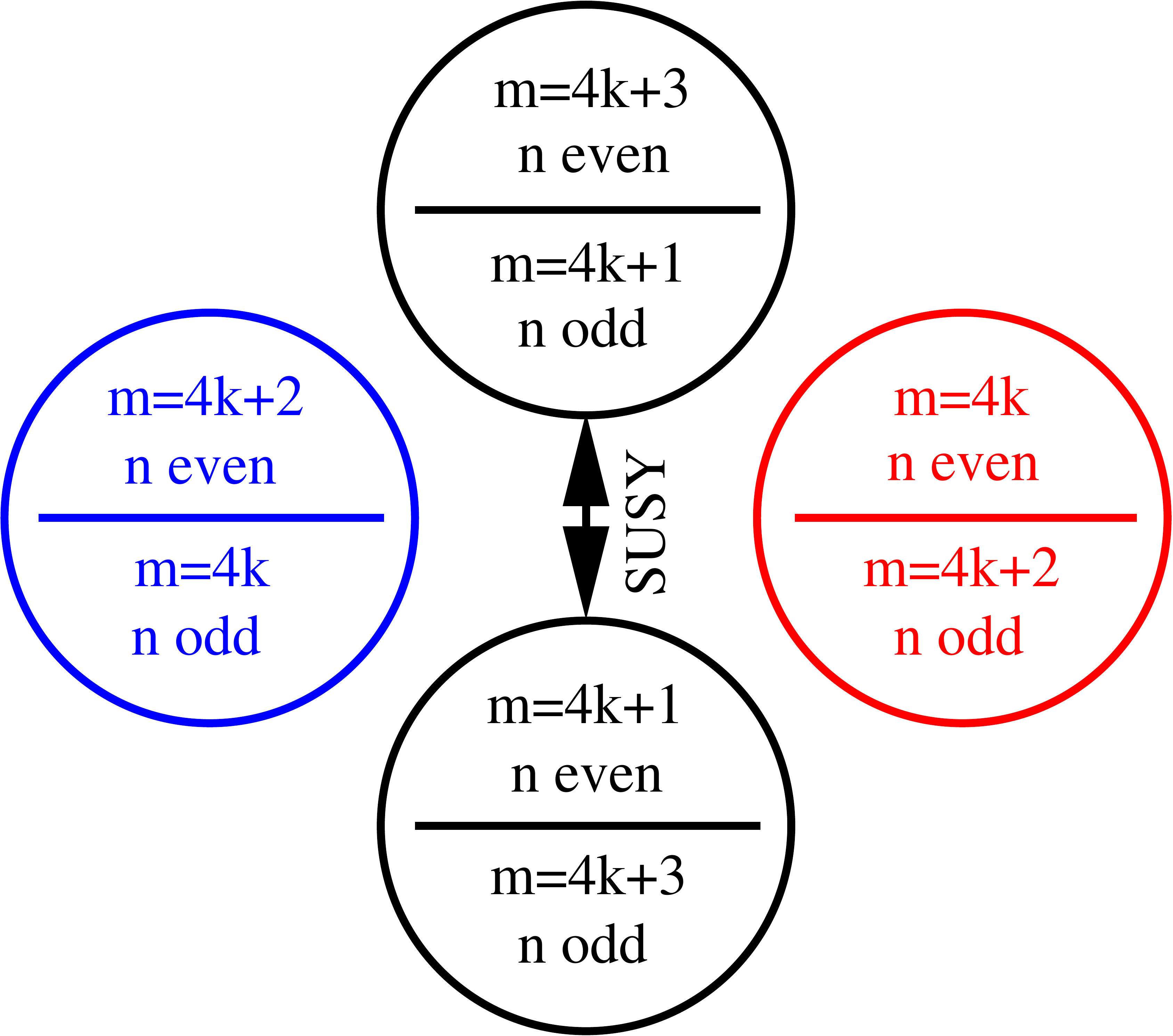}
\caption{Graphical representation of the four sets of product
  basis states spanning $\mathcal{H}_\ell, \ell=0,1,2,3$
  (clockwise from 3 o'clock) according to their eigenvalue
  with respect to the symmetry transform $\op{U}$, see
  \fmref{E-004B}. $m$ denotes the magnetic 
  quantum number, $k$ is an integer, and $n$ is the oscillator
  quantum number.}
\label{spin-phonon-susy-4}
\end{figure}

The case of a \emph{linear coupling}, where the unexpected level
crossings for odd integer spins occur, needs a deeper
investigation.
The key to understanding this phenomenon is provided by an
underlying not yet considered supersymmetry together with some
reasonable estimates. To this end we rewrite Hamiltonian
\fmref{E-000} 
using for the normal mode the coordinate operator
$\op{\xi}\propto
\left(\op{a}^{\dagger}+\op{a}^{\mathstrut}\right)$  
\begin{eqnarray}
  \op{H}_{\text{SI}}
  &=&
  D \big(\op{s}_z\big)^2
  +
  \alpha \sqrt{2\mu\omega}\ \op{\xi} \left[\big(\op{s}_x\big)^2-\big(\op{s}_y\big)^2\right]
  \ ,
  \label{E-003}
\end{eqnarray}
with $\mu$ being the oscillator mass ($\hbar=1$ throughout the
paper). 
It is now more obvious that this operator, and also the full
Hamiltonian without Zeeman term, have got a fourfold symmetry
with respect to the following symmetry operation 
\begin{eqnarray}
  \op{U} &=& \exp\left\{- i \pi \op{s}_z/2 \right\}\otimes \op{\Pi}
  \ ,
  \label{E-004}
\end{eqnarray}
which inverts $\xi$ (parity operation $\op{\Pi}$ acting on
$\xi$) and simultaneously rotates the spin vector operator about
its $z$-axis by $\pi/2$. The cyclic group generated by $\op{U}$
is of order four and has got four irreducible representations
that may be labeled by their characters $\exp\{-i \pi \ell/2\},
\ell=0,1,2,3$. All four irreps are realized by product
basis states that are already eigenstates of $\op{U}$,
\begin{eqnarray}
  \op{U} \ket{m, n}
  &=&
  \exp\left\{- i \pi m/2 \right\} (-1)^n
  \ket{m, n}
  \label{E-004B}
  \\
  &=&
  (-i)^m (-1)^n \ket{m, n}
  \ ,
  \label{E-004C}
\end{eqnarray}
and can thus be grouped according to these eigenvalues.
Therefore, the total Hilbert space can be decomposed into four mutually
orthogonal subspaces $\mathcal{H}=\mathcal{H}_0 \oplus
\mathcal{H}_1 \oplus \mathcal{H}_2 \oplus \mathcal{H}_3$. 
This is graphically depicted in \figref{spin-phonon-susy-4}.

The system possesses a second symmetry 
\begin{eqnarray}
  \op{V} &=& \exp\left\{- i \pi \op{s}_x \right\}\otimes \EinsOp
  \ ,
  \label{E-005}
\end{eqnarray}
which affects the spin part only. It leaves $\op{s}_x$ invariant
and rotates $\op{s}_y$ and $\op{s}_z$ into their respective
negatives. This operation also commutes with the Hamiltonian,
since $\op{H}$ depends on the squares of these operators. But
symmetry $\op{V}$ does not commute with $\op{U}$, at least not
on the full Hilbert space. However, thanks to the properties of
Hamiltonian \fmref{E-000B}, basis states $\ket{m, n}$
are only connected to $\ket{m, n}$ by $(\op{s}_z)^2$ and 
$\ket{m\pm 2, n}$ by $(\op{s}^+)^2$ and $(\op{s}^-)^2$, which
divides the Hilbert space into a direct sum 
of two mutually orthogonal subspaces for even and odd $m$, i.e.
\begin{eqnarray}
  \mathcal{H}&=&\mathcal{H}_{\text{even}}\oplus\mathcal{H}_{\text{odd}}
  \ ,\quad \text{with}
  \label{E-006A}
  \\
  \mathcal{H}_{\text{even}} &=& \mathcal{H}_0 \oplus
  \mathcal{H}_2
  \ ,
  \label{E-006B}
  \\
  \mathcal{H}_{\text{odd}} &=& \mathcal{H}_1 \oplus
  \mathcal{H}_3
  \label{E-006C}
  \ .  
\end{eqnarray}
$\op{U}$ and $\op{V}$ commute on $\mathcal{H}_{\text{even}}$,
whereas they anticommute on $\mathcal{H}_{\text{odd}}$.
Using concepts from supersymmetry, where $\op{U}$ and $\op{V}$
can be embedded into a Lie superalgebra \cite{Kac:77,Sou:PD85},
one can derive the following conclusions, see also supplemental 
material. 

One can show that symmetry $\op{V}$ maps $\mathcal{H}_1$ onto
$\mathcal{H}_3$ and \emph{vice versa} and eigenstates of $\op{H}$ that
are element of one of these two subspaces onto the respective
eigenstates in the
other space. Therefore, their energy eigenvalues must be at
least twofold degenerate. $\op{V}$ leaves $\mathcal{H}_0$ and
$\mathcal{H}_2$ invariant. Since $\mathcal{H}_1$ and 
$\mathcal{H}_3$ contain the states with odd $m$ quantum number,
these states are bound to be degenerate at $B=0$ and thus have
to cross.
Eigenstates with even values of $m$ are not degenerate by
symmetry, except for a possible, but unlikely accidental
degeneracy. These levels split, and therefore we 
observe an avoided level crossing in such cases.

Although from the point of view of applications only interesting
for the ground state, this observation holds also for excited
states. All levels that have been degenerate for $E=0$ split
under linear coupling to phonons if $m$ is an even integer and
they remain degenerate if $m$ is an odd integer.

The question whether the pair of levels that make up the
ground state without anisotropy, i.e.\ without coupling to the phonons, remains a
(tunnel-split) pair of ground state levels shall be answered
using perturbation theory.
If the interaction with the phonon subsystem is weak, i.e.\ much
weaker than given by the energy scale provided by the easy-axis
anisotropy $D$ -- and only these systems are technologically
interesting -- we find that the ground states consist to a large
extent of 
\begin{eqnarray}
  \ket{m=s, n=0}\ \text{and}\ \ket{m=-s, n=0} 
  \label{E-007}
\end{eqnarray}
for odd $m$ and of the two superpositions
\begin{eqnarray}
  \ket{m=s, n=0} \pm \ket{m=-s, n=0} 
  \label{E-008}
\end{eqnarray}
for even $m$. Admixtures of other basis states remain very
small, see analytical examples for $s=1$ and $s=2$ in the
supplemental material. Therefore, the related energy
eigenvalues of the ground states also deviate only little from those
of the axial system with $E=0$. Our numerical studies
for spin quantum numbers up to $s=8$ and $n_{\text{max}}=5$,
of which a part is shown in 
Figs.~\xref{spin-phonon-susy-2} and \xref{spin-phonon-susy-3},
arrive at the same conclusions.

\emph{Discussion.}---Our findings provide an interesting insight
into the effect of certain phonon modes on the tunneling gap at an avoided
level crossing. Counter-intuitive for a physics approach, the
linear term of a power series describing the interaction of the
phonon mode with the $E$ term of the anisotropy tensor -- which one
would naively assume to have the strongest effect -- does not lead to
any tunnel splitting in the case of odd integer spin quantum
numbers. It is the quadratic term that does the job.

The symmetry argument we found holds for all odd powers
of $\left(\op{a}^{\dagger}+\op{a}\right)$ where no splitting is
observed for odd integer spins, whereas for all even powers
thereof, a tunnel splitting exists.

Further on, the argument carries
through also for coupled spins. If spins interact via a
Heisenberg interaction, and if 
the phonons affect the anisotropy tensors as described, our
findings hold for the zero-field split multiplets in case of
total integer spin.

Thus, we understand from a more fundamental point of view why a
phonon that tilts an anisotropy tensor, as investigated in 
\cite{IrS:PRB20}, always opens a tunneling gap (for any
integer spin). The tilt, expressed as changes of both $E$ and $D$,
yields a Taylor series in $E$ that contains only even powers of
the oscillator elongation. Thanks to the zero-point motion of
the oscillator, this leads to $E>0$ and an immediate opening of
the tunneling gap, as explained above. In addition, also $D$ is
modified contrary to the investigation in this paper.

In a real magnetic molecule many phonon modes contribute to the
physical behavior
\cite{Shr:PR69,Shr:PLA70,Shr:CPL73,VMB:PSS78,PBK:PRL02,RBM:NC16,ESG:JPCL17,MST:NC18,ORD:DT19}.
However, the rational design of ligands and chemical bonds aims
at reducing the number of decohering and relaxing low-energy
modes. It is therefore desirable qualitatively understand the
character of the remaining active modes. With this article we
hope to contribute insight for a class of supersymmetric
spin-phonon systems where due to an odd-even effect the impact
on the tunneling gap is known \emph{a priori}.

Odd-even effects appear in many places in physics. In the context
of tunneling and supersymmetry, we found an article on
\emph{Inelastic cotunneling into a superconductor nano particle},
where odd and even numbers of tunneling electrons behaved
differently \cite{Aga:S99}, for curiosity.

\emph{Acknowledgment.}---This work was supported by the Deutsche
Forschungsgemeinschaft DFG 
355031190 (FOR~2692); 397300368 (SCHN~615/25-1)). We thank Eugenio
Coronado, Luis Escalera-Moreno, Stefano Sanvito, Mihail
Atanasov, Nick Chilton as well as Mark Pederson for valuable
discussions.

\emph{Corresponding author.}---J{\"urgen} Schnack, ORCID 0000-0003-0702-2723.


%

\end{document}


\title{Supplentary material: Supersymmetric spin-phonon coupling prevents odd integer spins from quantum tunneling}

\author{Kilian Irl\"ander}
\affiliation{Fakult\"at f\"ur Physik, Universit\"at Bielefeld, Postfach 100131, D-33501 Bielefeld, Germany}
\author{Heinz-J\"urgen Schmidt}
\affiliation{Universit\"at Osnabr\"uck, Fachbereich Physik,
 	D-49069 Osnabr\"uck, Germany}
\author{J\"urgen Schnack}
\email{jschnack@uni-bielefeld.de}
\affiliation{Fakult\"at f\"ur Physik, Universit\"at Bielefeld, Postfach 100131, D-33501 Bielefeld, Germany}

\date{\today}

\maketitle

In this supplementary material, we will substantiate the statements in the main part 
of the paper about the role of supersymmetry for the problem under consideration, see Section
\ref{sec:AS}. In the following two sections, we will further confirm the numerical results
on the degeneracy of the ground state by exact analytical diagonalization of the Hamiltonian
(without Zeeman term) for the case of $s=1$, see Section \ref{sec:ED}, and by
rigorous estimates for the analogous case of $s=2$, see Section  \ref{sec:GS}.

\section{Role of supersymmetry}\label{sec:AS}
The even-odd effect described in the main part of the paper depends on the fact
that the second symmetry
\begin{equation}\label{AS1}
  \op{V}=\exp\left\{ - i \pi \op{s}_x\right\}\otimes\op{\mathbbm 1}
\end{equation}
permutes the eigenspaces ${\mathcal H}_\ell,\;\ell=0,1,2,3$, of the first symmetry
\begin{equation}\label{AS2}
  \op{U}=\exp\left\{- i \pi \op{s}_z/2 \right\}\otimes \op{\Pi}
  \;,
\end{equation}
in the following way:
\begin{eqnarray}\label{AS3}
&&  \op{V}\,{\mathcal H}_0={\mathcal H}_0,\; \op{V}\,{\mathcal
    H}_2={\mathcal H}_2,
  \\
  \nonumber
&\mbox{but }&
   \op{V}\,{\mathcal H}_1={\mathcal H}_3\; \mbox{and }\op{V}\,{\mathcal H}_3={\mathcal H}_1
   \;.
\end{eqnarray}
This in turn follows from the observation that $\op{U}$ and $\op{V}$ commute on the subspace
${\mathcal H}_{\text{even}}={\mathcal H}_0 \oplus {\mathcal H}_2$, but {\em anti}-commute on 
${\mathcal H}_{\text{odd}}={\mathcal H}_1 \oplus {\mathcal H}_3$.
In fact, let $\phi\in  {\mathcal H}_\ell,\,\ell=1,3,$ be an eigenvector of $\op{U}$,
\begin{equation}\label{AS4}
 \op{U}\,\phi= (-i)^\ell\,\phi
 \;,
\end{equation}
then it follows that
\begin{eqnarray}
\label{AS5a}
  \op{U}\left( \op{V}\,\phi\right)&=&- \op{V}\,\op{U}\,\phi \\
  \label{AS5b}
   &\stackrel{(\ref{AS4})}{=}&-\op{V} \, (-i)^\ell\,\phi\\
   \label{AS5c}
   &=&  (-i)^{\ell+2}\,\left(\op{V}\,\phi\right)
   \;.
\end{eqnarray}
Hence $\op{V}\phi\in {\mathcal H}_{\ell+2}$, where $\ell+2$ is understood modulo $4$, and
$\op{V}$ maps ${\mathcal H}_\ell$ onto ${\mathcal H}_{\ell+2}$ for $\ell=1,3,$
if $\left\{\op{U},\op{V}\right\}=0$ on ${\mathcal H}_{\text{odd}}$.

This particular situation concerning the symmetries $\op{U},\,\op{V}$ can be
conveniently reformulated by using concepts from supersymmetry. This reformulation
could also be useful to identify other examples that fit into the same scheme.
In particular, we will embed $\op{U}$ and $\op{V}$ into a Lie superalgebra
such that these symmetries ``super-anticommute" in a sense to be explained.
Since the spacial factors of  $\op{U}$ and $\op{V}$, namely
$\op{\Pi}$ and $\op{\mathbbm 1}$, 
commute anyway, it will suffice to consider their spin factors and hence
to confine ourselves to finite-dimensional Lie superagebras.

Recall that a Lie superalgebra (LSA) is defined as a ${\mathbb Z}_2$-graded
algebra. This means it is a
linear space (over the field ${\mathbb R}$ or ${\mathbb C}$) of the form
\begin{equation}\label{AS6}
 {\mathcal  A} =  {\mathcal  A}_0 \oplus    {\mathcal  A}_1
 \;,
\end{equation}
equipped  with a bilinear map
\begin{equation}\label{AS7}
 [\,,\,\}: {\mathcal  A}\times {\mathcal  A}\rightarrow {\mathcal  A}
 \;,
\end{equation}
called the ``super-bracket". An element $a\in {\mathcal A}_0$ or $a\in {\mathcal A}_1$ is called ``homogeneous
of degree $|a|$" if $a\in {\mathcal A}_{|a|}$ and the following axioms (\ref{AS8a}) - (\ref{AS8c})
are understood to hold for homogeneous elements.
\begin{eqnarray} \nonumber
  |\,\left[a , b\right\}\,|& =&\left| a\right| + \left| b \right| \; \mod 2\;,\\
  \label{AS8a}
 && \quad \mbox{``${\mathbb Z}_2$-grading"}\\
 \nonumber
  \left[ a , b \right\} &=&-\,(-1)^{|a|\,|b|}  \left[ b , a\right\}\;,\\
 \label{AS8b}  
 && \quad \mbox{``(anti)symmetry"}\\
 \nonumber
\left[ a ,\left[  b ,c\right\} \right\}&=&
 \left[\left[a,b \right\} ,c \right\}+
(-1)^{|a|\,|b|}
\left[ b ,\left[  a ,c \right\}\right\}
\;, \\
\label{AS8c}
&&\mbox{``Jacobi identity"}
\end{eqnarray}
see, e.~g., \cite{Kac:77}.

In the following we will only use a special complex  LSA defined as follows:
Let $s$ be an (integer) spin quantum number and ${\mathcal M}_0$ denote the space
of all complex\\ $(2s+1)\times(2s+1)$-matrices. Let ${\mathcal M}_1$ be a copy of
 ${\mathcal M}_0$ such that
 \begin{equation}\label{AS9}
  {\mathcal M}={\mathcal M}_0 \oplus {\mathcal M}_1
  \;.
 \end{equation}
 The matrices $M\in{\mathcal M}_0$ will be called ``even" and those of ${\mathcal M}_1$ will be called ``odd".
 The Lie superbracket
 $\left[\right.\, , \, \left.\right\} $
 is defined as the commutator $[A,B]\in {\mathcal M}_0$
 for $A,B\in {\mathcal M}_0$, or, similarly, as $[A,B]\in {\mathcal M}_1$ for $A\in {\mathcal M}_0$ and $B\in {\mathcal M}_1$.
 On the other hand, the superbracket between two odd matrices $A,B\in {\mathcal M}_1$ is defined as the
 anti-commutator $\{ A,B\}\in  {\mathcal M}_0$. Finally, the superbracket is extended to ${\mathcal M}$
 by means of bilinearity. It is straightforward to show that (\ref{AS8a}) - (\ref{AS8c}) is satisfied and
 hence $\left({\mathcal M},\left[ ,\right\}\right)$ will be a complex LSA.

 We will denote by ${\sf U}$ and ${\sf V}$ the spin factors of $\op{U}$ and $\op{V}$, resp.~, that
 w.~r.~t.~the eigenbasis $|m\rangle,\;m=-s,\ldots,s,$  of $\op{s}_z$ assume the form
\begin{equation}\label{AS10}
 {\sf U}=\left(
\begin{array}{ccccccccc}
 \ddots & 0 & 0 & 0 & 0 & 0 & 0 & 0 & 0 \\
 0 & -i & 0 & 0 & 0 & 0 & 0 & 0 & 0 \\
 0 & 0 & -1 & 0 & 0 & 0 & 0 & 0 & 0 \\
 0 & 0 & 0 & i & 0 & 0 & 0 & 0 & 0 \\
 0 & 0 & 0 & 0 & 1 & 0 & 0 & 0 & 0 \\
 0 & 0 & 0 & 0 & 0 & -i & 0 & 0 & 0 \\
 0 & 0 & 0 & 0 & 0 & 0 & -1 & 0 & 0 \\
 0 & 0 & 0 & 0 & 0 & 0 & 0 & i & 0 \\
 0 & 0 & 0 & 0 & 0 & 0 & 0 & 0 & \ddots \\
\end{array}
\right)
\;,
\end{equation}
and
\begin{equation}\label{AS11}
 {\sf V}=(-1)^s\left(
\begin{array}{ccccccccc}
 0 & 0 & 0 & 0 & 0 & 0 & 0 & 0 & .\cdot{}^{\cdot} \\
 0 & 0 & 0 & 0 & 0 & 0 & 0 & 1 & 0 \\
 0 & 0 & 0 & 0 & 0 & 0 & 1 & 0 & 0 \\
 0 & 0 & 0 & 0 & 0 & 1 & 0 & 0 & 0 \\
 0 & 0 & 0 & 0 & 1 & 0 & 0 & 0 & 0 \\
 0 & 0 & 0 & 1 & 0 & 0 & 0 & 0 & 0 \\
 0 & 0 & 1 & 0 & 0 & 0 & 0 & 0 & 0 \\
 0 & 1 & 0 & 0 & 0 & 0 & 0 & 0 & 0 \\
 .\cdot{}^{\cdot} & 0 & 0 & 0 & 0 & 0 & 0 & 0 & 0 \\
\end{array}
\right)
\;.
\end{equation}
Next, we split ${\sf U}$ and  ${\sf V}$ into ``even" and ``odd" parts according to
\begin{eqnarray}
\label{AS12a}
 {\sf U} &=& {\sf U}_0 +{\sf U}_1 \\
 \nonumber
   &=& \left(
\begin{array}{ccccccccc}
 \ddots & 0 & 0 & 0 & 0 & 0 & 0 & 0 & 0 \\
 0 & 0 & 0 & 0 & 0 & 0 & 0 & 0 & 0 \\
 0 & 0 & -1 & 0 & 0 & 0 & 0 & 0 & 0 \\
 0 & 0 & 0 & 0 & 0 & 0 & 0 & 0 & 0 \\
 0 & 0 & 0 & 0 & 1 & 0 & 0 & 0 & 0 \\
 0 & 0 & 0 & 0 & 0 & 0 & 0 & 0 & 0 \\
 0 & 0 & 0 & 0 & 0 & 0 & -1 & 0 & 0 \\
 0 & 0 & 0 & 0 & 0 & 0 & 0 & 0 & 0 \\
 0 & 0 & 0 & 0 & 0 & 0 & 0 & 0 & \ddots \\
\end{array}
\right)\\
\label{AS12b}
& + &
\left(
\begin{array}{ccccccccc}
 \ddots & 0 & 0 & 0 & 0 & 0 & 0 & 0 & 0 \\
 0 & -i & 0 & 0 & 0 & 0 & 0 & 0 & 0 \\
 0 & 0 & 0 & 0 & 0 & 0 & 0 & 0 & 0 \\
 0 & 0 & 0 & i & 0 & 0 & 0 & 0 & 0 \\
 0 & 0 & 0 & 0 & 0 & 0 & 0 & 0 & 0 \\
 0 & 0 & 0 & 0 & 0 & -i & 0 & 0 & 0 \\
 0 & 0 & 0 & 0 & 0 & 0 & 0 & 0 & 0 \\
 0 & 0 & 0 & 0 & 0 & 0 & 0 & i & 0 \\
 0 & 0 & 0 & 0 & 0 & 0 & 0 & 0 & \ddots \\
\end{array}
\right)
\;,
\end{eqnarray}
and
\begin{eqnarray}
\label{AS13a}
 {\sf V} &=& {\sf V}_0 +{\sf V}_1 \\
 \nonumber
   &=&(-1)^s \left(
\begin{array}{ccccccccc}
 0 & 0 & 0 & 0 & 0 & 0 & 0 & 0 & .\cdot{}^{\cdot} \\
 0 & 0 & 0 & 0 & 0 & 0 & 0 & 0 & 0 \\
 0 & 0 & 0 & 0 & 0 & 0 & 1 & 0 & 0 \\
 0 & 0 & 0 & 0 & 0 & 0 & 0 & 0 & 0 \\
 0 & 0 & 0 & 0 & 1 & 0 & 0 & 0 & 0 \\
 0 & 0 & 0 & 0 & 0 & 0 & 0 & 0 & 0 \\
 0 & 0 & 1 & 0 & 0 & 0 & 0 & 0 & 0 \\
 0 & 0 & 0 & 0 & 0 & 0 & 0 & 0 & 0 \\
 .\cdot{}^{\cdot} & 0 & 0 & 0 & 0 & 0 & 0 & 0 & 0 \\
\end{array}
\right)\\
\label{AS13ab}
&+&(-1)^s
\left(
\begin{array}{ccccccccc}
 0 & 0 & 0 & 0 & 0 & 0 & 0 & 0 & .\cdot{}^{\cdot} \\
 0 & 0 & 0 & 0 & 0 & 0 & 0 & 1 & 0 \\
 0 & 0 & 0 & 0 & 0 & 0 & 0 & 0 & 0 \\
 0 & 0 & 0 & 0 & 0 & 1 & 0 & 0 & 0 \\
 0 & 0 & 0 & 0 & 0 & 0 & 0 & 0 & 0 \\
 0 & 0 & 0 & 1 & 0 & 0 & 0 & 0 & 0 \\
 0 & 0 & 0 & 0 & 0 & 0 & 0 & 0 & 0 \\
 0 & 1 & 0 & 0 & 0 & 0 & 0 & 0 & 0 \\
 .\cdot{}^{\cdot} & 0 & 0 & 0 & 0 & 0 & 0 & 0 & 0 \\
\end{array}
\right)
\;.
\end{eqnarray}
We embed these parts into ${\mathcal M}$ such that
\begin{equation}\label{AS14}
 {\sf U}_0,\,{\sf V}_0\in {\mathcal M}_0\;\mbox{and  } {\sf U}_1,\,{\sf V}_1\in {\mathcal M}_1
 \;.
\end{equation}
W.~r.~t.~this embedding it can be easily shown that the following holds:
\begin{prop}\label{P1}
 \begin{equation}\label{AS15}
   \left[ {\sf U},  {\sf V}  \right\}=0
   \;,
 \end{equation}
 that is,
 \begin{eqnarray}\label{AS16}
 && [{\sf U}_0,{\sf V}_0]= [{\sf U}_0,{\sf V}_1]= [{\sf V}_0,{\sf
      U}_1]=0,
  \\
 &&  \mbox{and }
  \{{\sf U}_1,{\sf V}_1\}=0
\nonumber
  \ .
 \end{eqnarray}
\end{prop}

Thus the general scenario where we can expect a similar alternating behaviour between level crossing and
avoided level crossing as in the present paper is the case where there exist two super-commuting symmetries.
Further aspects of super-symmetric quantum mechanics like the occurrence of super-symmetric pairs
of Hamiltonians do not appear to be realized in the present case.

\section{Exact diagonalization for $s=1$}\label{sec:ED}

The phenomenon of degenerate eigenspaces caused by
super-symmetry occurs for all Hamiltonians of the form provided
by Eqs.~(4) and (8) in the main document, 
such that $B=0$ and hence $\op{H}_{\text{Zeeman}}=0$. The latter condition will be tacitly assumed in the remainder of this Section,
i.~e., we always write
\begin{eqnarray}\label{ED1a}
  \op{H}&=& \op{H}_{\text{SI}} + \op{H}_{\text{HO}}\\
  \nonumber
  &=&
   D (\op{s}_z)^2 \otimes \op{\mathbbm 1}+
  \frac{\alpha}{2} \left\{ (\op{s}^+)^2+({\op{s}^-})^2 \right\}\otimes \op{x}\\
  \label{ED1b}
 && +  \op{\mathbbm 1}\otimes \omega \left(\op{a}^{\dagger}\op{a}^{\mathstrut}+\frac{1}{2}\right)
  \;,
\end{eqnarray}
using $ \left( (\op{s}_x)^2-(\op{s}_y)^2 \right)=\frac{1}{2} \left( (\op{s}^+)^2+({\op{s}^-})^2 \right)$.
Additionally, the question arises whether the ground state is degenerate, i.e., whether the ground state lies in one of the subspaces
${\mathcal H}_1$ or ${\mathcal H}_3$ and hence in both. Numerical evidence suggests that this will be the case for $D<0$ and odd $s$.
In this section we will confirm this finding by exact diagonalization of the Hamiltonian for $s=1$ and $B=0$.

Since the Hamiltonian $\op{H}$ leaves the eigenspaces ${\mathcal H}_\ell$ of $\op{U}$ for $\ell=0,1,2,3$ invariant,
it is possible to perform the diagonalization for each of the four subspaces separately. Due to the symmetry $\op{V}$
only one of the two cases ${\mathcal H}_1$ or ${\mathcal H}_3$ needs to be considered.

\subsection{Subspace ${\mathcal H}_1$}\label{sec:EDI}

The subspace ${\mathcal H}_1$ is spanned by the product states
\begin{equation}\label{EDI1}
 |m,n\rangle=|(-1)^{n+1},n\rangle,\,n=0,1,2,\ldots
 \;.
\end{equation}
Let ${\sf H}$ denote the matrix of $\op{H}$ w.~r.~t.~this basis. It is tri-diagonal
since $\op{x}$ is represented by a tri-diagonal matrix. The diagonal elements of ${\sf H}$
are obtained as
\begin{eqnarray}\label{EDI2a}
{\sf H}_{nn}&=&
  \left\langle (-1)^{n+1},n \left|\, \op{H}\, \right|(-1)^{n+1},n\right\rangle\\
  \nonumber
 &=&  D\, \left\langle (-1)^{n+1} \left|(\op{s}_z )^2 \right|(-1)^{n+1}\right\rangle\\
 \label{EDI2b}
&&  +
   \omega \,\left\langle n \left|\left(\op{a}^{\dagger}\op{a}^{\mathstrut}+{\textstyle\frac{1}{2}}\right)\right|n\right\rangle\\
   \label{EDI2c}
   &=& D+\omega \left(n+{\textstyle\frac{1}{2}}\right)
   \;,
\end{eqnarray}
since $\langle n| \op{x}|n\rangle=0$. For the upper secondary diagonal of ${\sf H}$ we have
\begin{eqnarray}
\label{EDI3a}
  {\sf H}_{n,n+1} &=&  \left\langle (-1)^{n+1},n \left|\, \op{H}\, \right|(-1)^{n},n+1\right\rangle\\
  \nonumber
   &=&\frac{\alpha}{2}  \left\langle (-1)^{n+1} \left| (\op{s}^+)^2+({\op{s}^-})^2 \right|(-1)^{n}\right\rangle\, \times\\
   &&
   \label{EDI3b}
    \left\langle n \left| \op{x} \right|n+1\right\rangle\\
 \label{EDI3c}
 &=& \alpha\,\sqrt{2\,\mu\,\omega\,n}
 \;,
\end{eqnarray}
since the secondary diagonal matrix elements of $({s}_z)^2$ and $\op{H}_{\text{HO}}$ vanish
and
\begin{equation}\label{EDImat}
  (\op{s}^+)^2+({\op{s}^-})^2=
  \left(
\begin{array}{ccc}
 0 & 0 & 2 \\
 0 & 0 & 0 \\
 2 & 0 & 0 \\
\end{array}
\right)
\;.
\end{equation}
Obviously, ${\sf H}_{n+1,n}={\sf H}_{n,n+1}$.

It follows that ${\sf H}$ is the same matrix as that of the
operator
\begin{equation}\label{EDI4}
 \op{K}\equiv \op{H}_{\text{HO}}+D\,\op{\mathbbm 1}+\alpha\,\op{x}
 \;,
\end{equation}
w.~r.~t.~the harmonic oscillator eigenbasis $|n\rangle,\;n=0,1,2,\ldots$.
Neglecting for the moment the constant energy shift due to $D\,\op{\mathbbm 1}$ we may
write
\begin{eqnarray}
\label{EDI5a}
    \op{H}_{\text{HO}}+\alpha\,\op{x} &=& \frac{1}{2\mu}\op{p}^2+\frac{\mu\omega^2}{2}\op{x}^2+\alpha\op{x} \\
   \nonumber
    &=& \frac{1}{2\mu}\op{p}^2+\frac{\mu\omega^2}{2}\left(\op{x} +x_0\right)^2-\frac{\mu\omega^2}{2}x_0^2
    \;,\\
     \label{EDI5b}
    &&
 \end{eqnarray}
 where
 \begin{equation}\label{EDI6}
  x_0\equiv \frac{\alpha}{\mu\omega^2}
  \;.
 \end{equation}
We conclude that ${\sf K}$ is the matrix of a harmonic oscillator Hamiltonian with a spatially shifted minimum of the potential and
a constant energy shift of
\begin{equation}\label{EDI8}
 \delta E=D-\frac{\mu\,\omega^2}{2}x_0^2 \stackrel{(\ref{EDI6})}{=} D-\frac{\alpha^2}{2\,\mu\,\omega^2}
 \;.
\end{equation}
Its eigenvalues are hence of the form
\begin{equation}\label{EDI9}
  E^{(1)}_n=\omega\left( n+{\textstyle \frac{1}{2}}\right) + D-\frac{\alpha^2}{2\,\mu\,\omega^2}
  \;,
\end{equation}
with the relative ground state energy
\begin{equation}\label{EDI10}
 E^{(1)}_0=\frac{\omega}{2} + D-\frac{\alpha^2}{2\,\mu\,\omega^2}
 \;.
\end{equation}
Moreover, this result supports the remark in the main document
referring to Eq.~(15),
since the ground state of the shifted oscillator and the ground
state of the  unshifted have a considerable overlap for small
$x_0$.

\subsection{Subspace ${\mathcal H}_0$}\label{sec:EDO}

The subspace ${\mathcal H}_0$ is spanned by the product states
\begin{equation}\label{EDO1}
 |m,n\rangle=|0,n\rangle,\,n=0,2,4,\ldots
 \;.
\end{equation}
Since the matrix elements of $\op{x}$ between different states of this basis vanish
the matrix ${\sf K}$ of the restriction of $\op{H}$ to the subspace ${\mathcal H}_0$ is already
of diagonal form. Its diagonal elements that represent the energy eigenvalues read
\begin{eqnarray}\label{EDO2}
{\sf K}_{n,n}&=&  \left\langle 0,n\left|{\mathbbm 1}\otimes \op{H}_{\text{HO}}
+D(\op{s}_z )^2\otimes {\mathbbm 1}\right| 0,n\right\rangle\\
&=& \omega\,\left( n+{\textstyle \frac{1}{2}}\right)
\;,
\end{eqnarray}
for $n=0,2,4,\ldots$, and yield the relative ground state energy
\begin{equation}\label{EDO3}
 E^{(0)}_0=\frac{\omega}{2}
 \;.
\end{equation}

\subsection{Subspace ${\mathcal H}_2$}\label{sec:EDII}

Analogously to Subsection \ref{sec:EDO}, the subspace ${\mathcal H}_2$ is spanned by the product states
\begin{equation}\label{EDII1}
 |m,n\rangle=|0,n\rangle,\,n=1,3,5,\ldots
 \;.
\end{equation}
Since the matrix elements of $\op{x}$ between different states of this basis vanish,
the matrix ${\sf K}$ of the restriction of $\op{H}$ to the subspace ${\mathcal H}_0$ is already
of diagonal form. Its diagonal elements read
\begin{eqnarray}\label{EDII1B}
{\sf K}_{n,n}&=&  \left\langle 0,n\left|{\mathbbm 1}\otimes \op{H}_{\text{HO}}
+D (\op{s}_z )^2\otimes {\mathbbm 1}\right| 0,n\right\rangle\\
&=& \omega\,\left( n+{\textstyle \frac{1}{2}}\right)
\;,
\end{eqnarray}
for $n=1,3,5,\ldots$, and yield the relative ground state energy
\begin{equation}\label{EDII10}
 E^{(2)}_1=\frac{3\,\omega}{2}\stackrel{(\ref{EDO3})}{>}E^{(0)}_0
 \;.
\end{equation}

\subsection{Total ground state}\label{sec:TGS}
Summarizing the results of the Subsections \ref{sec:EDI} - \ref{sec:EDII} we conclude that
$E_0^{(1)}$ according to (\ref{EDI10}) represents the total ground state energy since
\begin{equation}\label{TGS1}
  E_0^{(1)}-E_0^{(0)}= D-\frac{\alpha^2}{2\,m\,\omega^2}<0
  \;
\end{equation}
according to the assumption $D<0$ made in this Section. This completes the arguments for the
groundstate lying in the subspace ${\mathcal H}_{\text{odd}}$ in the case of $s=1$.

\section{Ground state for $s=2$}\label{sec:GS}

According to numerical evidence, the groundstate lies in ${\mathcal H}_{\text{even}}$ for even $s$.
We will confirm this result by rigorous estimates of the (relative) ground state energies for $s=2$.
We again consider the Hamiltonian (\ref{ED1b}) and the various invariant subspaces ${\mathcal H}_\ell,\;\ell=0,1,2,3$.

\subsection{Subspace ${\mathcal H}_1$}\label{sec:GSI}

The results of Subsection \ref{sec:EDI} can largely be adopted, with the exception that for $s=2$ 
the matrix of\\ $(\op{s}^+)^2+({\op{s}^-})^2$ assumes the form
\begin{equation}\label{GSI1}
 (\op{s}^+)^2+({\op{s}^-})^2=
 \left(
\begin{array}{ccccc}
 0 & 0 & 2 \sqrt{6} & 0 & 0 \\
 0 & 0 & 0 & 6 & 0 \\
 2 \sqrt{6} & 0 & 0 & 0 & 2 \sqrt{6} \\
 0 & 6 & 0 & 0 & 0 \\
 0 & 0 & 2 \sqrt{6} & 0 & 0 \\
\end{array}
\right)
\;.
\end{equation}
By comparison with (\ref{EDImat}) this means that the parameter $\alpha$ in Subsection \ref{sec:EDI} 
has to be replaced by $3\,\alpha$ for the present case which gives the new expressions for
the eigenvalues
\begin{equation}\label{GSI2}
 E^{(1)}_n= \omega\left(n+{\textstyle \frac{1}{2}}\right) +D -\frac{9 \, \alpha^2}{2\,\mu\,\omega^2}
 \;,\quad \mbox{for } n=0,1,2\ldots
 \end{equation}
 and the relative ground state energy
 \begin{equation}\label{GSI3}
  E^{(1)}_0= \frac{\omega}{2} +D -\frac{9 \, \alpha^2}{2\,\mu\,\omega^2}
  \;.
 \end{equation}

\subsection{Subspace ${\mathcal H}_2$}\label{sec:GSII}

The subspace ${\mathcal H}_2$ is spanned by the product states
\begin{equation}\label{GSII1}
 |m,n\rangle = |\pm 2,0\rangle, |0,1\rangle, |\pm 2,2\rangle, |0,3\rangle, \ldots
 \;.
\end{equation}
It can be further split into the two eigenspaces of $\op{V}$ formed by 
symmetric or antisymmetric linear combinations of the states $|\pm m,n\rangle$ in (\ref{GSII1}).
These eigenspaces are also left invariant by the Hamiltonian $\op{H}$.
In the following, we only consider the symmetric subspace
${\mathcal H}_{2,s}$ 
spanned by the states
\begin{equation}\label{GSII2}
|\tilde{2},0\rangle, |0,1\rangle, |\tilde{2},2\rangle, |0,3\rangle, \ldots
\;,
\end{equation}
where $ |\tilde{2}\rangle$ denotes the spin state
\begin{equation}\label{GSII3}
 |\tilde{2}\rangle=\frac{1}{\sqrt{2}}\left( |2\rangle +|-2\rangle\right)
 \;.
\end{equation}

Let ${\sf K}$ denote the matrix of the Hamiltonian  $\op{H}$ w.~r.~t.~the basis (\ref{GSII2}).
Its diagonal entries read
\begin{eqnarray}\label{GSII4a}
 {\sf K}_{n,n}&=&\left\langle \tilde{2},n \left| \,{\mathbbm 1}\otimes \op{H}_{\text{HO}}
 + D\,( \op{s}_z)^2 \otimes {\mathbbm 1}\, \right|  \tilde{2}, n\right\rangle\\
 \label{GSII4b}
 &=& \omega\left( n+{\textstyle \frac{1}{2}}\right) + 4 D,\quad \mbox{for even } n
 \;.
\end{eqnarray}
Here we have used that the state $|\tilde{2}\rangle$ is an eigenstate of $( \op{s}_z)^2$ 
corresponding to the eigenvalue $m^2=4$. Analogously,
\begin{eqnarray}\label{GSII5a}
 {\sf K}_{n,n}&=&\left\langle 0,n \left| \,{\mathbbm 1}\otimes \op{H}_{\text{HO}}
 + D\,( \op{s}_z)^2 \otimes {\mathbbm 1}\, \right|  0, n\right\rangle\\
 \label{GSII5b}
 &=& \omega\left( n+{\textstyle \frac{1}{2}}\right) ,\quad \mbox{for odd } n
 \;,
\end{eqnarray}
since the state $|0\rangle$ is an eigenstate of $( \op{s}_z)^2$
corresponding to the eigenvalue $m^2=0$.

For the upper secondary diagonal entries of ${\sf K}$, we first consider the case of even $n$ and obtain
\begin{eqnarray}
  \nonumber
    {\sf K}_{n,n+1}
    &=&\frac{\alpha}{2} \left\langle 0,n \left| 
   \,\left((\op{s}^+)^2+({\op{s}^-})^2\right)\otimes \op{x}
   \,\right| \tilde{2},n+1\right\rangle
   \\
    &=& \frac{\alpha}{2\sqrt{2}}\left(
   \langle 0\left|(\op{s}^+)^2+({\op{s}^-})^2 \right| 2\rangle\right.
   \label{GSII6a}
   \\
  && +\left.
   \langle 0\left|(\op{s}^+)^2+({\op{s}^-})^2 \right|- 2\rangle
   \right)
   \langle n| \op{x}|n+1\rangle
   \nonumber
   \\
   \label{GSII6b}
   &\stackrel{(\ref{GSI1})}{=}&   \frac{\alpha}{2\sqrt{2}} \left( 2\times 2\sqrt{6}\right) \sqrt{2\,\mu\,\omega\, n}\\
   &=&\alpha \,2\sqrt{3} \, \sqrt{2\,\mu\,\omega\, n}
   \;.
\end{eqnarray}
The result for odd $n$ is the same. It follows that, analogously to Subsection \ref{sec:EDI},
${\sf K}$ equals the matrix of a shifted harmonic oscillator Hamiltonian $\widehat{\op{H}}$ plus
a diagonal operator $\op{\Delta}$. Here,
\begin{equation}\label{GSII7}
  \widehat{\op{H}}=\frac{1}{2\mu}\op{p}^2+\frac{\mu\omega}{2}\left(\op{x}+x_0\right)^2-\frac{\mu\omega}{2}x_0^2
  \;,
\end{equation}
where
\begin{equation}\label{GSII8}
  x_0=\frac{2\sqrt{3}\alpha}{\mu\omega^2}
  \;,
\end{equation}
and hence
\begin{equation}\label{GSII9}
  \widehat{\op{H}}=\frac{1}{2\mu}\op{p}^2+\frac{\mu\omega}{2}\left(\op{x}+x_0\right)^2-\frac{6\alpha^2}{\mu\omega^2}
  \;.
\end{equation}
Further,
\begin{equation}\label{GSII10}
 \op{\Delta}=4\,D\,\mbox{diag}\left( 1,0,1,0,\ldots,\right)
 \;.
\end{equation}

We want to determine an upper bound of $E^{(2)}_0$ of the form
\begin{equation}\label{GSII11}
E^{(2)}_0\le \left\langle \Phi\left|  \widehat{\op{H}}+\op{\Delta} \right| \Phi\right\rangle
\;,
\end{equation}
where $\Phi$ is chosen as the normalized ground state of $ \widehat{\op{H}}$, to wit,
\begin{equation}\label{GSII12}
  \Phi(x)= \left(\frac{\mu\omega}{\pi}\right)^{1/4}\, \exp\left(-\frac{\mu\omega}{2}\left(x+x_0\right)^2 \right)
  \;.
\end{equation}
Further, we will have to use the explicit form of the harmonic oscillator eigenfunctions
\begin{eqnarray}
\nonumber
  \phi_n(x) &=& \left(2^n\,n!\,\sqrt{\frac{\pi}{\mu\omega}}
  \right)^{-1/2} \\
  \label{GSII13}
   && \exp \left(-\frac{\mu\omega}{2} x^2\right)\,H_n\left(\sqrt{\mu\omega}x\right)
   \;,
\end{eqnarray}
where $H_n(\ldots)$ denotes the $n$-th Hermite polynomial.

We first note that
\begin{equation}\label{GSII14}
 \left\langle \Phi\left|  \widehat{\op{H}} \right| \Phi\right\rangle=\frac{\omega}{2}-\frac{6\alpha^2}{\mu\omega^2}
 \;.
\end{equation}
Then we consider
\begin{equation}\label{GSII15}
 \left\langle \Phi\left|  \op{\Delta}\right| \Phi\right\rangle=4\,D\,\sum_{n=0,2,\ldots}
 \langle \Phi|n\rangle\,\langle n| \Phi\rangle
 \;.
\end{equation}
After some calculations we obtain the intermediate result
  \begin{eqnarray}
 \nonumber
  \langle \Phi|n\rangle &=& \int_{-\infty}^{\infty} \Phi(x)\,\phi_n(x)\,dx\\
     \label{GSII16}
  &=& \frac{\left(\sqrt{\mu\omega x_0} \right)^n}{\sqrt{2^n n!}}\,\exp\left( 
  -\frac{\mu\omega x_0^2}{4}
  \right)
  \;.
\end{eqnarray}
and hence
\begin{eqnarray}
\label{GSII17a}
 \left\langle \Phi\left|  \op{\Delta}\right| \Phi\right\rangle
 &=&
 4\,D\,\sum_{n=0,2,\ldots} \left|  \langle
 \Phi|n\rangle\right|^2
 \\
 &=& 4\,D\,\exp\left( -\frac{\mu\omega x_0^2}{2} \right)
 \cosh\left(\frac{\mu\omega x_0^2}{2} \right)
 \nonumber
 \\
 &=&
 2\,D\,\left( 1+\exp\left( -\mu\omega x_0^2 \right)\right)
\nonumber
 \\
 \label{GSII17d}
 &\stackrel{(\ref{GSII8})}{=}&2\,D\,\left( 1+\exp\left( -\frac{12\alpha^2}{\mu\omega^3} \right)\right)
 \;.
\end{eqnarray}
Summarizing,
\begin{eqnarray}
\nonumber
 E_0^{(2)} &\le& \frac{\omega}{2}-\frac{6\alpha^2}{\mu\omega^2} \\
 \label{GSII18}
 &&+2\,D\,\left( 
 1+\exp\left(
 -\frac{12 \alpha^2}{\mu\omega^3}
 \right)
 \right)
 \;.
\end{eqnarray}
\\

\subsection{Total ground state}\label{sec:GST}
Combining the previous results, we obtain
\begin{eqnarray}
 E_0^{(2)}-E_0^{(1)} &\stackrel{(\ref{GSII18},\ref{GSI3})}{\le}&-\frac{3}{2}\frac{\alpha^2}{\mu\omega^2}\\
 \label{GST1a}
\nonumber
  &&+ D\left( 1+2 \exp\left( -\frac{12 \alpha^2}{\mu\omega^3} \right) \right)\\
  \label{GST1b}
  & <& 0
  \;,
\end{eqnarray}
since $D<0$. This proves that $E_0^{(1)}$ cannot be the total ground state energy and hence the total
ground state cannot lie in ${\mathcal H}_{\text{odd}}$.
Actually, the numerical calculations show that it lies in the subspace ${\mathcal H}_{2,s}$ considered above, 
while the above analytical considerations prove only the weaker
result that the total ground state lies in ${\mathcal
  H}_{\text{even}}$. 

Finally, we would like to provide an example for our statement
that the ground state with spin-phonon coupling consists mainly
of the two ground states for $E=0$. The ground state in the
discussed case of $s=2$ ($D=-5$, $n_{\text{max}}=1$, 
  $\alpha=0.5$, and $\omega=5$ in natural units) is
\begin{eqnarray}
  \ket{\Phi_0} &=&
  +0.706684 \ket{m=2, n=0}
\\
&&+ 0.706684\ket{m=-2, n=0}
\nonumber
\\
&&- 0.034579\ket{m=0, n=1}
\nonumber
  \ .
\end{eqnarray}
Thus, it contains only 0.1~\% admixture of other states.


%